\documentclass{article} 
\usepackage{iclr2022_conference,times}

\usepackage{amsmath,amsfonts,bm}

\def\eqref#1{equation~\ref{#1}}

\def\1{\bm{1}}

\DeclareMathAlphabet{\mathsfit}{\encodingdefault}{\sfdefault}{m}{sl}
\SetMathAlphabet{\mathsfit}{bold}{\encodingdefault}{\sfdefault}{bx}{n}

\usepackage{graphicx}
\usepackage{hyperref}
\usepackage{url}

\title{A Brief Guide to Designing and \\ Evaluating Human-Centered \\ Interactive Machine Learning}

\author{Kory W. Mathewson \& Patrick M. Pilarski \\ 
DeepMind \\
\texttt{\{korymath,ppilarski\}@deepmind.com} \\
}

\iclrfinalcopy 
\begin{document}

\maketitle

\begin{abstract}
Interactive machine learning (IML) is a field of research that explores how to leverage both human and computational abilities in decision making systems. IML represents a collaboration between multiple complementary human and machine intelligent systems working as a team, each with their own unique abilities and limitations. This teamwork might mean that both systems take actions at the same time, or in sequence. Two major open research questions in the field of IML are: ``How should we design systems that can learn to make better decisions over time with human interaction?'' and ``How should we evaluate the design and deployment of such systems?'' A lack of appropriate consideration for the humans involved can lead to problematic system behaviour, and issues of fairness, accountability, and transparency. Thus, our goal with this work is to present a human-centred guide to designing and evaluating IML systems while mitigating risks. This guide is intended to be used by machine learning practitioners who are responsible for the health, safety, and well-being of interacting humans. An obligation of responsibility for public interaction means acting with integrity, honesty, fairness, and abiding by applicable legal statutes. With these values and principles in mind, we as a machine learning research community can better achieve goals of augmenting human skills and abilities. This practical guide therefore aims to support many of the responsible decisions necessary throughout the iterative design, development, and dissemination of IML systems.
\end{abstract}

\section{Introduction and Background}
\label{sec:introbg}

\begin{figure}[ht!]
    \centering
    \includegraphics[width=\textwidth]{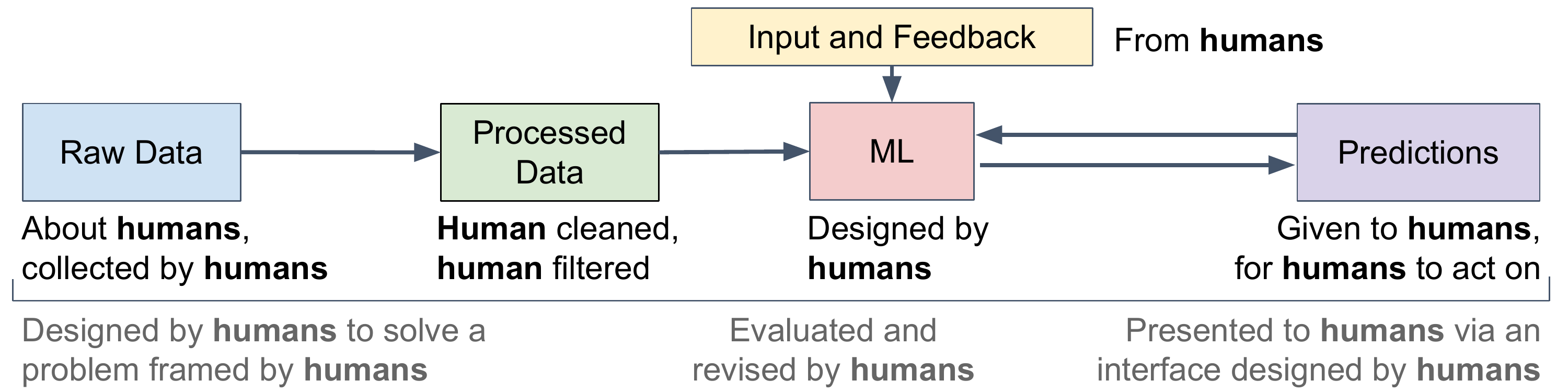}
    \caption{All machine learning is interactive machine learning as there are humans in the loop. The core of the ML system is designed by humans, who also provide raw or processed inputs to the system, as well as labels and feedback in the form of demonstrations, corrections, rankings, evaluations, advice, or guidance. IML systems often present humans with predictions for the humans to then act on. Interaction can be loosely or tightly coupled and can happen before, during, or after training. Interaction between humans and ML systems can be considered sociotechnical interaction.}
    \label{fig:human}
\end{figure}

Machine learning (ML) comprises a set of computing science techniques for automating knowledge acquisition rather than relying on explicit human instruction. 
Interactive ML (IML) systems interface with humans to incorporate human input, activity, assistance, demonstrations, feedback, and, more generally, knowledge and information.
Perhaps surprisingly, \textit{all ML systems can be considered to have humans in their learning loop} (see Figure \ref{fig:human}). While some interactions are more tightly coupled with humans than others, human interaction is a key element in all ML. For example, interacting humans may provide data, objective functions, direct feedback, algorithms, or code. All of this can be considered IML. An IML system might incorporate multiple modes of interaction. 
To date, designers have demonstrated a range of behaviours in how they acknowledge and take responsibility for the human interaction that is key to IML. 

Evaluating IML systems from a human-centred perspective requires re-framing the process to include humans during conceptualization, implementation, and interaction. Human-centred thinking can be well expected to lead to more usable ML systems, improved mutual understanding, and augmented communication between intelligent systems~\citep{pilarski2017communicative}. These benefits come at the increased cost of time and energy to understand human factors. As a result, some choose to ignore social factors in their pursuit of IML. This choice has serious implications, including issues of fairness and accountability and misalignment with human values~\citep{char2018implementing, zafar2017fairness,leike2018scalable}.

Human-centred design involves developing solutions to problems by involving the human perspective in all steps of the process---such systems put people before machines~\citep{cooley1996human}. While ML traces its history back decades~\citep{samuel1967some}, it was not until \citet{Fails:2003:IML:604045.604056} that the term interactive machine learning (IML) was introduced. More recently, \cite{amershi2014power} and \cite{aaai2017tut} provide a detailed review the field and the role of humans in IML systems.
Research incorporating human decision making with ML systems is gaining momentum as many realize the importance of human-centered thinking in the design and deployment of increasingly capable ML technologies~\citep{DBLP:journals/corr/MathewsonP16,ibarz2018reward,leike2018scalable,DBLP:journals/corr/abs-1709-03969}. Previous work has discussed the virtues of human-centered thinking in ML and the mechanisms for accountability especially related to datasets~\citep{lovejoy2017human,10.1145/3442188.3445918,gebru2021datasheets}, including the importance of participatory design~\citep{berditchevskaia2021Participatory}. But, there continues to be a gap between the mental models of responsible practitioners and many contemporary deployments.


This guide details many of the questions that we as a community would likely want addressed to support the claims made in IML research papers. This is critical to ML evaluation, as an increasing number of ML systems begin to find their way out of laboratory settings without standard institutional ethics review \citep{mckee2022inreview} due to their lack of alignment with the kinds of settings in which review is typically required. For instance, even though many ML studies will have no human subjects, research outputs such as code, data, models and predictions are released and used by the general public. Design, development, and deployment can happen without subscribing human subjects directly.

Sec. \ref{sec:hcd} focuses on how we should design IML systems that can learn to make better decisions over time with human interaction. This section also presents two foresight exercises, which connect design thinking to human-centered engineering~\citep{Hehn2020OnID}--which can help streamline iterations, and a \textit{premortem}--which can help identify and mitigate potential risks and failure modes early in the process. Sec. \ref{sec:dev} concerns itself with the iterative development and evaluation process---a reflexive process of looking back and reviewing to make progress. It advocates for checking in with the humans-in-the-loop through each iteration. Sec. \ref{sec:diss} focuses on questions of model deployment and public communication. We hope that it will prove helpful to read through this guide before starting an IML project and to continually check-in with it throughout design, development, and dissemination of human-interactive decision-making systems.

\section{Human-Centered Design}
\label{sec:hcd}

\paragraph{Step 1: Define the hypothesis.} Good design starts with clarity. Clearly state your investigated question of interest. Can you pose it as a testable hypothesis \citep{popper2005logic}, which can be supported by evidence? This premise ought to motivate investigation and support the responsible and ethical use of the human interaction effort required for IML. 

\paragraph{Step 2: Loop in humans.} Define your values and principles. How will you align this work with human needs? Identify individual and societal social factors in the problem of interest. Start by considering why your hypothesis is essential not just to you, but also to the larger community of people who might interact with it. How will the system make people's lives better? Consider your hypothesis from multiple stakeholder perspectives. As an example, think about three groups of individuals: those who might be invested in your work, those who might be impacted by your work, and those who might be interested in your work. Empathizing with stakeholders helps you to appreciate how the problem, and your potential solutions, will affect them. Leverage participatory design by involving stakeholders in ideation. Commence impact assessments: what are the environmental, safety, privacy, and human-rights implications of your work?

Consent starts with communication. Build open-communication channels with stakeholders in these groups. Gather ideas, design requirements, concerns, and questions from them. You should review your values alongside the values of these individuals. How will these stakeholders engage with your system? Thinking about this now will help during deployment in Step 9. How might the interaction look? Ask each of \textit{them} how \textit{they} will evaluate the performance of your system? Discuss how the system might be used both constructively and misapplied to harm. This dual-use discussion is ongoing in the field of ML~\citep{gpt2}. Choices you make will impact these people directly, and you are responsible for the impact of your work on them. Embrace this responsibility. These stakeholders can champion your system if you engage them early in the process and often through iteration~\citep{ganguli2022predictability}.

\paragraph{Step 3: Define the goal.} Define a specific, measurable, attainable, relevant, and timely goal. It should be linked directly to the hypothesis from Step 1. The goal should clearly define success for the humans involved. This definition will ideally encompass all the ways that your stakeholders will engage with, and evaluate the performance of, your system. There are often multiple metrics which define success for a given problem. ML system designers typically refine optimization to a single metric of interest. Consider both your optimization metric (i.e. model performance indicator) and your measures of system success for humans interacting with your system. Ethical research must consider the benefit to society and to individuals, and the risks to both. The benefit of the people involved in the interaction is a factor in ML system success, as is the impact on people who engage with people who are interacting with the ML technology. How does this learning objective align with the ways that your stakeholders will evaluate your system? Define a testing suite for safety which you will use in Step 6. These tests should consider human health, safety, and well-being. As well, they should evaluate your system on biases, fairness, and equality across hidden features in your data \citep{10.1145/3287560.3287596,DBLP:journals/corr/abs-1711-09883}.

It can help to align your work with familiar categories of existing ML work \citep{langley2000crafting}. Is this project developing a new model, applying existing methods to new data, or presenting a new model of human behaviour? Does the system test the limitations of current models on a new problem? Given your goal, what are the technical, scientific, implementation problems which need to be solved? You should be able to break your system into components (e.g. data, processing, evaluation). Doing so ought to make addressing each part individually easier. What are some of the downstream impacts of accomplishing your goal? That is, if the results support your hypothesis, what else might be true?

\paragraph{Step 4: Define the data.} An ML system often reflects the training data it learns on~\citep{10.1145/3287560.3287596}. It can reflect many common human biases, and thus you ought to consider dataset coverage and organization. What is your ideal data set? How much data do you desire? How much data do you need? Why might these amounts be different? What are the dependent and independent variables? How will the data be organized and represented for the learning system? How might possible data sources stray from the ideal data? How will you define what outliers and bad data points are?

Data engineering and processing questions require consideration. How will you accumulate, clean, parse, label, and safely store your data? How might you fill in blanks in your data? Can you use software to simulate data? Once you collect your data, split it into training, evaluation (i.e. validation), and held-out testing data segments. Do this early, lest you leak information from the test segment into your model selection and parameter tuning processes. Experiments on simulated data designed to test assumptions and gain intuition provide valuable insights. How will you incorporate new data which arrives after deployment?

You should consider dataset collection, annotation, ownership, and privacy. How will you handle participant recruitment and compensation? If you pay for data (e.g. through crowd-sourcing, direct payment to humans, or a third party), what are the costs of accumulating data (and how does compensation align with basic living wage in the region of data collection)? What are the usage rights and responsibilities of your data? What is the ownership model for this data? What mechanisms are in place for data to be withdrawn from your system?

If you have humans in your data collection, consider the ethical implications of collecting their data. How are your data generating humans informed of the use of their data? How are data privacy and security communicated? What are the potential biases and sensitivities in human-collected data (i.e. personal or identifying information)? How will you secure formal, rigorous, arms-length ethics reviews and approvals for protocols that use human participation and human data?

\subsection{Design Thinking Exercises}

\paragraph{The Whiteboard Model.}
For the whiteboard model exercise, consider the following: given your hypothesis, stakeholder analysis, and goal, how would you get to a solution given a short amount of time and only a whiteboard? While it is tempting to think about novel techniques and solutions which might address your goal, it is often more effective to make something that works and then make iterative improvements. This thought exercise will also provide an opportunity to mentally zoom-out from the problem and think about how potential solutions fit in the system as a whole.

\paragraph{The Premortem.}
For the premortem exercise, imagine that the project fails for a variety of reasons. Write down these failure modes. Then, for each failure mode, work backwards to identify what might have lead to different results. This process of prospective hindsight can increase the ability to identify the reasons for future outcomes by 30\%   \citep{mitchell1989back,klein_2007}. Finally, working forward, assess how histories of related designs might also unfold within or influence the human interactions in the present project. A premortem can provide insights and ideas which you can use in the next iterative development steps and can help reduce the chances of arriving at predictable failure modes.

\section{Develop, Analyze, Evaluate, and Iterate}
\label{sec:dev}

\paragraph{Step 5: Build model.} Safe design is the first step towards safe use. Think about model misuse starting with the first model you build. Step 2 (Loop in Humans) covered much of this preparation.

Consider simple models for learning from your data. A simple model serves as a baseline for comparing model improvements. What is your baseline model? It might be a model that generates random outputs; a \textit{random} is a perfectly reasonable baseline and can help to identify other bugs in the development pipeline. Other reasonable benchmarks include a `majority-class' model that predicts the most common output in the training set and a `by-hand' model which invites a human to consider the inputs and generate an output.

The `by-hand' model is often called a Wizard-of-Oz, or human-assisted model, and has been used at scale to help understand human interactions~\citep{facebookM}. Another `by-hand' model is a pseudo-adversarial interaction by a human who is attempting to push the model toward poor performance. Recent work has also leveraged other models to generate adversarial test cases for your system~\citep{DBLP:journals/corr/abs-2202-03286}. These models can provide base-case and worst-case scenario performance on your metrics. Similar to the \textit{whiteboard model}, evaluating the performance of these models towards your goal will help to define essential features in your data, and in the larger system.

\paragraph{Step 6: Evaluate model.} Performance on your evaluation data will serve as a consistent comparison for model improvement. Test your model on your evaluation data segment. Track your key metrics. The performance of your baseline model starts as your `best', and `worst', performing model. Keep your model performances as comparisons as you iterate in Step 8 (Re-evaluate and Iterate). What are the limitations of your evaluation scheme? What are the unaccounted costs or errors? How does the model perform on the evaluation data and the safety suite designed in Step 3 (Define the goal).

With each evaluation iteration, it is essential to think about biases, fairness, and equality across diverse groups. Each iteration is a crucial checkpoint to communicate with stakeholders. Your stakeholders' discussions should include how they feel your model has addressed the ideas, interests, design requirements, concerns, and questions brought up in Step 2 (Loop in Humans). How do they evaluate system performance? How would the baseline model impact them? Have you considered the power dynamics between those that develop the model and those affected by its deployment?

\paragraph{Step 7: Analyze trade-offs.} You will make trade-offs as you iterate. Consider these trade-offs by listing each of them and their associated impacts independently. Trade-offs often include factors such as cost, storage, learning speed, inference speed, computation complexity, model serving, deployment, and human interpretability. It helps to perform ablation studies which systematically remove model components to determine their relative contributions. Considering each of these trade-offs will help you iterate on your model development.

\paragraph{Step 8: Re-evaluate and iterate} Given the trade-offs defined in Step 7, review your key metrics. Ensure you capture all the information required before continuing. For instance, how do you log experimental parameters, model information, and results? Once you are confident that you can systematically make model improvements towards your evaluation metric, then it is time to iterate through Steps 5, 6, and 7. Once your evaluation performance converges, only then should you test your model(s) on the held out test set data segment. This testing should be used to compare models, and not to tune model parameters.

\section{Disseminate}
\label{sec:diss}

\paragraph{Step 9: Deploy the system.} Present and test the system with your stakeholders and new individuals that you have not engaged with up to this point. When testing with humans, focus on usability. How are stakeholders interacting with your model? What do they say about your model and how do they say it? Have you considered qualitative measures of following usability during deployment, in addition to the usual quantitative measures? Usability can have a profound effect on the perceived quality and capability of models. These are valuable interactions, note how these humans interpret the performance of your system.

Consider that many humans may act against the system, by accident or on purpose. How will you handle attacks on your model? What are the technical implications of model security? What design choices have you made that influence the behaviour of those interacting with your model? What are the fail-safes and procedural safeguards? How can you adapt them during deployment? How are you communicating the risks of interaction?

The steps in this guide and foresight exercises can go a long way toward illuminating potential challenges and set backs with your work. But, no exercise can account for all potential harms, impacts, and outcomes. Thus, once deployed, how will you give people routes for system monitoring, complaint logging, withdrawal, and recourse?

\paragraph{Step 10: Communicate.} The purpose of communication is to convey the key ideas to your audience clearly so that they may comprehend them with minimal effort. You should be able to state your key results and how it aligns with your hypothesis. Do your results match or contradict similar work? What are the limitations of the current model? How might these problems be addressed in the future? Do the results challenge any of the ideas or beliefs of the stakeholders? When communicating the project, it is helpful to follow the ML Reproducibility checklist \citep{pineau2018repro}. Can you open source your code, data, models, and deployment? Consider how and why others might attempt reproduction.

\pagebreak
\section{Conclusions}
\label{sec:conc}

Human-centred and sociotechnical thinking for design and development can lead to substantial improvements in development and adoption. By empathizing with those invested in, impacted by, or adversarial of ML systems, developers can better serve the needs of all humans involved. Enabling humans to efficiently and effectively interact with systems continues to be a key design challenge~\citep{dudley2018review}. As highlighted through this guide, proactive thinking in designing and evaluating human-centered IML systems can help to address ongoing challenges of bias and unfairness. This has the potential to improve transparency and increase the accountability of those designing, developing and deploying IML systems.

\subsubsection*{Acknowledgments}
We thank the reviewers for their time and effort. We also thank Piotr Mirowski and Shakir Mohamed for comments on the paper, as well as William Isaac, Alex Zacherl, Laura Weidinger, and many others at DeepMind for the conversations have helped shape our views.

\bibliography{iclr2022_conference}
\bibliographystyle{iclr2022_conference}


\end{document}